\documentclass[manuscript]{aastex}

\def\'#1{\ifx#1i{\accent"13\i}\else{\accent"13#1}\fi}

\newcommand{\kms}{{\rm ~km~s}^{-1}}
\newcommand{\Mgas} {M_{\rm gas}}
\newcommand{\MJ} {M_{\rm J}}
\newcommand{\Msink} {M_{\rm sink}}
\newcommand{\Mstar} {M_*}
\newcommand{\Msun} {M_\odot}
\newcommand{\nth} {n_{\rm th}}
\newcommand{\pcc}{{\rm ~cm}^{-3}}
\newcommand{\psc}{{\rm ~cm}^{-2}}
\newcommand{\sfeff} {SFE$_{\rm ff}$}

\newcommand{\tff} {t_{\rm ff}}
\newcommand{\tsf} {t_{\rm SF}}
\newcommand{\tsfinv} {t_{\rm SF}^{-1}}
\newcommand{\VS}{V\'azquez-Semadeni}

\slugcomment{Submitted to ApJ}

\shorttitle{Star Formation from  Hierarchical Gravitational Fragmentation}
\shortauthors{V\'azquez-Semadeni et al.}

\begin{document}


\title{High- and Low-Mass Star Forming Regions from Hierarchical
Gravitational Fragmentation. High local Star Formation Rates with Low
Global Efficiencies} 


\author{Enrique V\'azquez-Semadeni\altaffilmark{1}, Gilberto
C. G\'omez\altaffilmark{1}, A.-Katharina Jappsen\altaffilmark{2},
Javier Ballesteros-Paredes\altaffilmark{1} and Ralf S. Klessen\altaffilmark{3}}


\altaffiltext{1}{Centro de Radioastronom\'\i a y Astrof\'\i sica,
Universidad Nacional Aut\'onoma de M\'exico, Apdo. Postal 3-72, Morelia,
Michoac\'an, 58089, M\'exico}
\altaffiltext{2}{School of Physics \& Astronomy, Cardiff University,
Queens Buildings, The Parade, Cardiff CF24 3AA, UK}
\altaffiltext{3}{Zentrum f\"ur Astronomie der Universit\"at Heidelberg,
Institut f\"ur Theoretische Astrophysik, 69120 Heidelberg, Germany}


\begin{abstract}
We investigate the properties of ``star forming regions'' in a
previously published numerical simulation of molecular cloud formation
out of compressive motions in the warm neutral atomic interstellar
medium, neglecting magnetic fields and stellar feedback. 
We study the
properties (density, total gas+stars mass, stellar mass, velocity
dispersion, and star formation rate) of the cloud hosting the first
local, isolated ``star formation'' event and compare them with those of
the cloud formed by the central, global collapse event. In this
simulation, the velocity dispersions at all scales are caused primarily
by infall motions rather than by random turbulence. We suggest that the
small-scale isolated collapses may be representative of low- to
intermediate-mass star-forming regions, with gas masses ($\Mgas$) of
hundreds of solar masses, velocity dispersions $\sigma_v
\sim 0.7 \kms$, and star formation rates (SFRs) $\sim 3 \times 10^{-5}
\Msun {\rm yr}^{-1}$, while the large-scale, massive ones may be
representative of massive star forming regions, with $\Mgas$ of
thousands of solar masses, $\sigma_v \sim$ a few $\kms$, and SFRs $\sim
3 \times 10^{-4} \Msun {\rm yr}^{-1}$. 
We also compare the statistical distributions of the physical properties
of the dense cores appearing in the central region of massive collapse
with those from a recent survey of the massive star
forming region in the Cygnus X molecular cloud, finding that the
observed and simulated distributions are in general very similar.

However, we find that the star formation efficiency per free-fall time
(\sfeff) of the high mass region, similarly to that of OMC-1, is low,
$\sim 0.04$. In the simulated cloud, this is {\it not} a consequence of
a ``slow'' SFR in a nearly hydrostatic cloud supported by turbulence,
but rather of the region accreting mass at a high rate. Thus, we find
that measuring a low \sfeff\ may be incorrectly interpreted as implying
a lifetime much longer than the core's local free-fall time, and an SFR
much slower than that given by the free-fall rate, if the accretion is
not accounted for. We suggest that, rather than requiring a low value of
the \sfeff\ everywhere in the Galaxy, attaining a globally low specific SFR
requires star formation to be a spatially intermittent process,
so that most of the mass in a GMC is not participating of the SF process
at any given time. Locally, the specific SFR of a star-forming region
can be much larger than the global GMC's average.

\end{abstract}



\section{Introduction} \label{sec:intro}

The formation of massive stars is currently a matter of intense
debate \citep[e.g.,][]{ZY07}. High-mass star forming regions are
characterized by more extreme 
physical conditions than their low-mass counterparts, 
containing cores of size, mass, and velocity dispersion roughly an
order of magnitude larger than those of cores in low-mass regions
\citep[e.g.][]{JMA99, LM99, GL99, Kurtz_etal00, Beuther_etal07}. In
particular, typical values of the properties of clumps within high-mass star
forming regions (HMRs) are sizes 0.2--0.5 pc, mean densities $n \sim
10^5$ cm$^{-3}$, masses between 100 and 1000 $M_\odot$, and velocity
dispersions ranging between 1.5 and 4 km s$^{-1}$. In turn, the clumps
break down into even denser ``cores'' that are believed to be the
immediate precursors of single or gravitationally bound multiple massive
protostars. 

The high velocity dispersions of clumps in HMRs are generally
interpreted as strong turbulence that manages to support the clumps
against gravity \citep[e.g.,][]{GL99,MT03}. However, the notion of
``turbulent support'' is difficult to maintain at the scales of these
cores. If turbulence is to ``support'' a cloud against self-gravity, it
must consist of isotropic motions, with typical scales significantly
smaller than the size of the cloud, so that collectively these motions
can act as a source of isotropic pressure, similarly to
thermal molecular motions \citep{MK04, BP_etal07}. 

However, turbulence is a flow regime in which
the largest velocity differences are associated with the largest
separations \citep[e.g.][]{Frisch95}, so that, in a turbulence-induced density
fluctuation (i.e., a cloud or clump), the largest velocity difference is
expected to occur at scales comparable to the size of the clump itself,
thus being {\it highly anisotropic} with respect to it. This property is
indeed observed in molecular clouds, in which a principal component
analysis of the velocity structure shows that the dominant mode is
dipole-like \citep{HeBr07}, indicating either shear or compression, but
not solid-body rotation (M.\ Heyer, 2009, private communication). Being
dipolar, the principal velocity component appears unsuited to act as a
supporting form of pressure capable of opposing self-gravity.

Moreover, in the case of supersonic turbulence, the clumps are expected
to be formed by large-scale compressive motions \citep{HF82, BVS99,
BKV03, Klessen_etal05}, suggesting that the turbulence {\it within the
clumps} is similarly likely to have a strong compressive
component. Indeed, \citet{VS_etal08} presented a numerical study of
turbulent clouds driving showing that, even with continuous {\it random}
driving at large scales, there is a trend for the clumps to contain a
net convergence of the flow. This result contradicts the hypothesis that
the non-thermal motions consist of isotropic, small-scale
turbulence. After all, the clumps {\it must} be formed by a globally
converging flow, as dictated by the continuity equation.

In addition, numerical simulations of cloud formation in the diffuse
atomic ISM \citep[e.g.][]{VPP95, VPP96, PVP95, BVS99, BHV99, KI02, AH05,
Heitsch_etal05, Heitsch_etal06, Heitsch_etal08, VS_etal06, VS_etal07,
HA07} and of star formation in turbulent, self-gravitating clouds
\citep[e.g.][]{KHM00, HMK01, Klessen01, BBB03, VKSB05, Jappsen_etal05,
VS_etal07, FKS08,
Federrath_etal09} show that the velocity fields are in 
general organized at all scales, exhibiting a continuity from the large
scales outside the clumps all the way to their interiors. Specifically,
the simulations of molecular cloud formation by \citet[][hereafter
Paper I]{VS_etal07}, showed that, as the cloud's mass is consumed by the
formation of collapsed objects (``stars''), the cloud continues to
accrete mass from its surrounding, more diffuse (``atomic'')
environment, the whole process amounting to a ``mass cascade'',
analogous to a turbulent energy cascade, as proposed by
\citet{NW90} and \citet{FBK08}. Observational evidence supporting this
scenario has been recently provided by \citet{Galvan_etal09}.

In this paper, we investigate the consequences of this hierarchical
fragmentation scenario on the formation of low- and high-mass star
forming regions, taking advantage of the variety of star forming regions
appearing in the main simulation of Paper I. This study was motivated by
a careful observation of the mechanism through which the clouds are
assembled in that simulation. Specifically, we show that
the region in the simulation with physical conditions resembling those
of observed massive star-forming regions forms as the result of the
large-scale collapse of a large complex, thus suggesting that such
regions may be in generalized gravitational collapse rather than in a
quasi-static state supported by turbulence, as proposed, for example, by
\citet{HB07} for the Orion molecular cloud.

The plan of the paper is as follows. In \S \ref{sec:model} we briefly
recall the main parameters and evolutionary features of the numerical
simulation, and identify the regions we investigate. In \S
\ref{sec:phys_cond_lo_hi} we measure the evolution of the total (dense
gas + ``stars'') and ``stellar'' masses, and the instantaneous star
formation rate (SFR) in both the isolated, peripheral, low-mass region
(LMR) and in the central, global-collapse HMR. In \S
\ref{sec:comparison_obs} we then compare our results with available
obervations. In \S \ref{sec:phys_props_SFRs} we compare the physical
conditions and the star formation rates, showing that these indicators
are reasonably similar to those of actual star forming regions of each
type. Next, in \S \ref{sec:core_stats_ctr} we perform a survey of the
cores in our HMR, and compare the distributions of their sizes,
masses and densities with those of a recent survey of the Cygnus X
region by \citet{Motte_etal07}, showing that the distributions are very
similar. Finally, in \S \ref{sec:conclusions} we 
discuss the limitations of our study and present a summary.

\section{The numerical model} \label{sec:model}


\subsection{Parameters} \label{sec:params}

In this paper, we analyze data from the simulation labeled L256$\Delta
v$0.17 in Paper I. We refer the reader to that paper for details. Here
we just mention that it is a smoothed particle hydrodynamics (SPH)
simulation, performed with the GADGET code \citep[][]{Springel_etal01},
including self-gravity and parameterized heating and cooling that imply
a thermally bistable medium.  The simulation uses $3.24 \times 10^6$ SPH
particles, and uses the prescription by \citet{Jappsen_etal05} for
forming ``sink particles'' at densities above a certain density
threshold, which we take as $n_{\rm thr} = 3.2 \times 10^6 \pcc$. Once
formed, the sinks can continue to accrete mass from their
surroundings. Note, however, that the sinks are not sufficiently
resolved in mass to be considered individual stars, and should be
considered stellar clusters instead.  The total mass in the simulation
is $5.26 \times 10^5 \Msun$, the size of the numerical box is 256 pc,
and the initial conditions are uniform, with $T = 5000$ K and a mean
density of $n = 1$ cm$^{-3}$, in which two cylindrical flows are set up
to collide at the center of the simulation.  The flows have the same
density and temperature as the rest of the box, but oppositely directed
velocities of $7.536 \kms$, corresponding to a Mach number of 1.22. The
flows have a length of 112 pc and a radius of 36 pc. 

It is worth noting that the cylindrical inflows are much narrower than
the length of the numerical box, having a diameter of only 1/4 of the
box length. This allows for interaction between the cloud and its
environment, and also frees the cloud from boundary effects. Also, the
length of the inflows is slightly shorter than the box length, so that
the inflows are entirely contained within the box. This is because the
boundaries are periodic, and thus the inflows cannot enter from the
outside. This implies also that the total mass in the box is fixed.

\subsection{Evolution} \label{sec:evolution}

The simulation we consider in this paper was designed to represent the
formation and initial evolution of a dense, cold cloud out of the
collision of streams of warm neutral gas. The collision nonlinearly
triggers a phase transition to the cold neutral phase \citep{HP99, WF00,
KI02, Heitsch_etal05, AH05, Heitsch_etal06, VS_etal06, VS_etal07, HA07},
forming a large, flattened, and cold atomic cloud, which is turbulent
because of a variety of instabilities in the interface between the cold
and warm components \citep{KI02, Heitsch_etal06, VS_etal06}. The
turbulence is moderately supersonic with respect to the cold gas.

Because the cloud is colder and denser, it readily becomes
self-gravitating and begins contracting and fragmenting, so that the
large atomic cloud becomes a large molecular-like cloud complex. The
densest fragments begin to undergo small-scale, local collapse events
\citep{Banerjee_etal09}. These isolated collapse events occur in a
peripheral ring, which forms because the collapse of a finite sheet-like
cloud proceeds from the outside in, as shown by \citet{BH04}. Also, the
local collapses occur {\it before} the bulk of the cloud completes its
global, large-scale collapse, and the clumps and the stars they form are
part of the global collapse of the complex. In the remainder of the
paper, we show that the site at which the global collapse finally
converges has physical conditions very similar to those of HMRs, while a
typical local-collapse region has physical conditions resembling those
of LMRs.

The large atomic cloud formed by the flow collision, which later becomes
a denser ``molecular cloud''\footnote{The simulation does not follow the
chemistry, but we refer to gas with density and temperature typical of
molecular gas as ``molecular''.} complex permeated by an atomic
substrate, has a flattened shape during most of its evolution. The
large-scale gravitational contraction begins at $t \sim 10$ Myr, and
star formation begins at $t \sim 17$ Myr. The initial radius of the
peripheral ring is $\sim 20$ pc. By $t = 23.4$ Myr, the global collapse
is completed (the ring has shrunk to nearly zero radius), although the
turbulence in the cloud has caused the infall motions to have a random
component, so that the collapse center spans several parsecs
across. Animations showing the large-scale evolution of this simulation
can be found in the electronic edition of Paper I.

It is important to note that in Paper I we applied a prescription by
\citet{FST94} to estimate {\it a posteriori} the time at which the total
stellar mass formed in the simulation would imply the presence of enough
massive stars to reionize the whole complex. We found this time to be
roughly 3 Myr after the onset of star formation, or roughly at $t \sim
20$ Myr. However, simulations including ionization-heating feedback
(\VS, Col\'in \& G\'omez, in prep.) suggest that, while the feedback is
capable of disrupting the small-scale clumps, it does not prevent the
large-scale collapse to form the massive region, since the material that
falls into the latter comes from much larger distances than the local
spheres of influence of the stellar sources \citep[see
also][]{Dale_etal05, DB08, Peters_etal09}. In addition, the cloud
continues accreting mass from the warm medium even while it is actively
forming stars. Thus, it is safe to assume that the
central massive region would form even if we had included stellar
feedback in the present simulation.

\subsection{The regions} \label{sec:the regions}

In this paper, we focus on two different regions of ``star'' (i.e., sink
particle) formation. As a representative example of an LMR, we consider
the first region to ever form stars in the peripheral ring starting at
$t = 17.3$ Myr, which involves relatively small mass in both gas and
sinks. Other LMRs in the simulation are similar to this one. As an
example of an HMR, we consider the central region, where the global
collapse converges at $t = 23.4$ Myr. We choose this time because it is
the time of maximum compression of the material (see the animations
corresponding to Fig.\ \ref{fig:cen8pc_images} in the electronic
edition). This region involves a substantially larger amount of mass
than the LMR, and is the only such region formed in the simulation,
because the total mass in the complex is only $\sim 5 \times 10^4~\Msun$
at the time we explore this region. Thus, this is our only choice for
this type of region. Presumably, a more massive simulation could form a
larger number of such regions.

\begin{figure}
\plotone{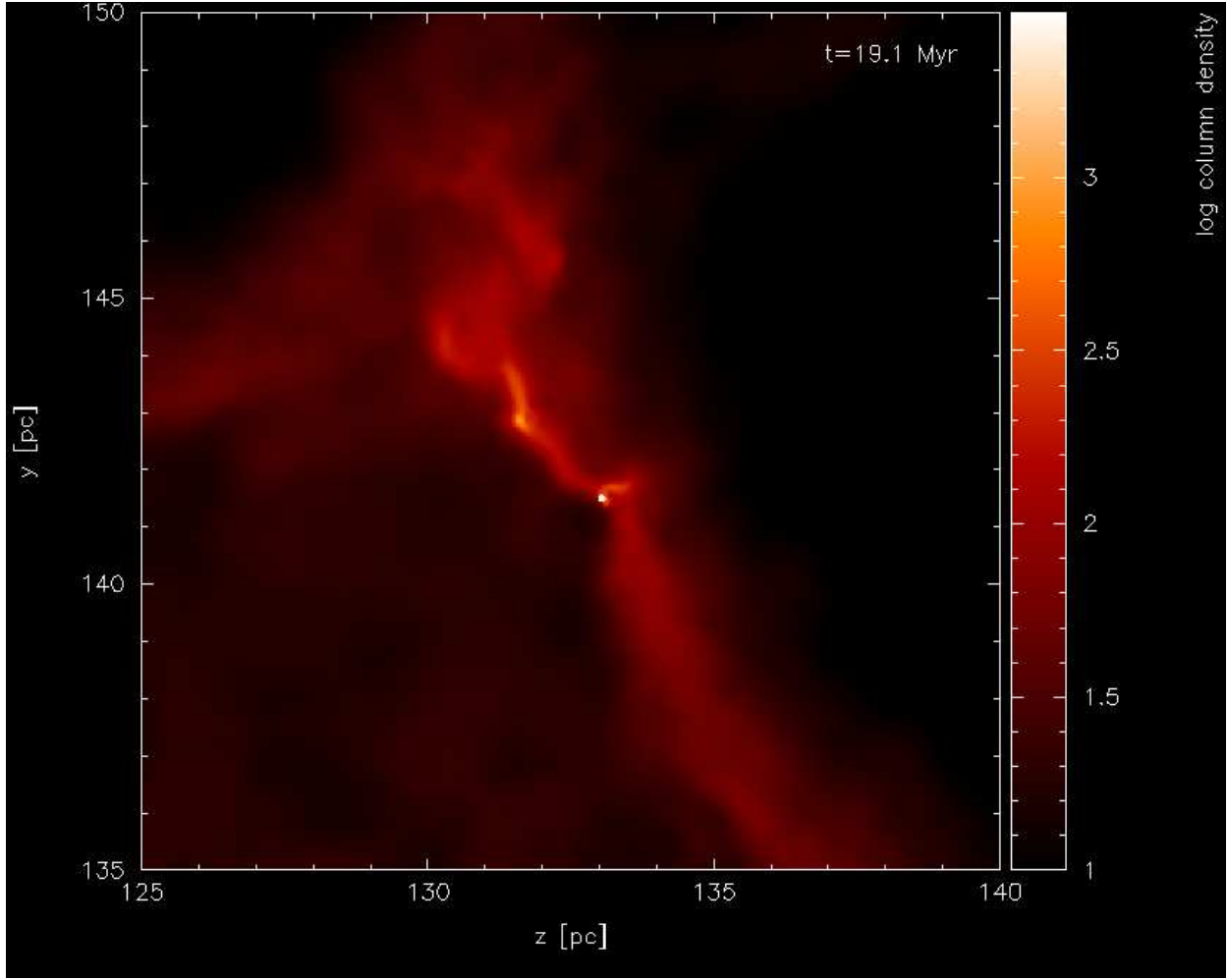}
\caption{Column density plot of Cloud 1 in the
$y$-$z$ plane at $t=19.1$ Myr, integrating over the central 16 pc along
the $x$ direction. The dots show the stellar objects (sink
particles). The electronic version of this figure shows an animation of
this region from $t=16.6$ to 19.9 Myr. The column density is in code
units, which correspond to $9.85 \times 10^{19} \psc$.}
\label{fig:core1_images}
\end{figure}
\begin{figure}
\plottwo{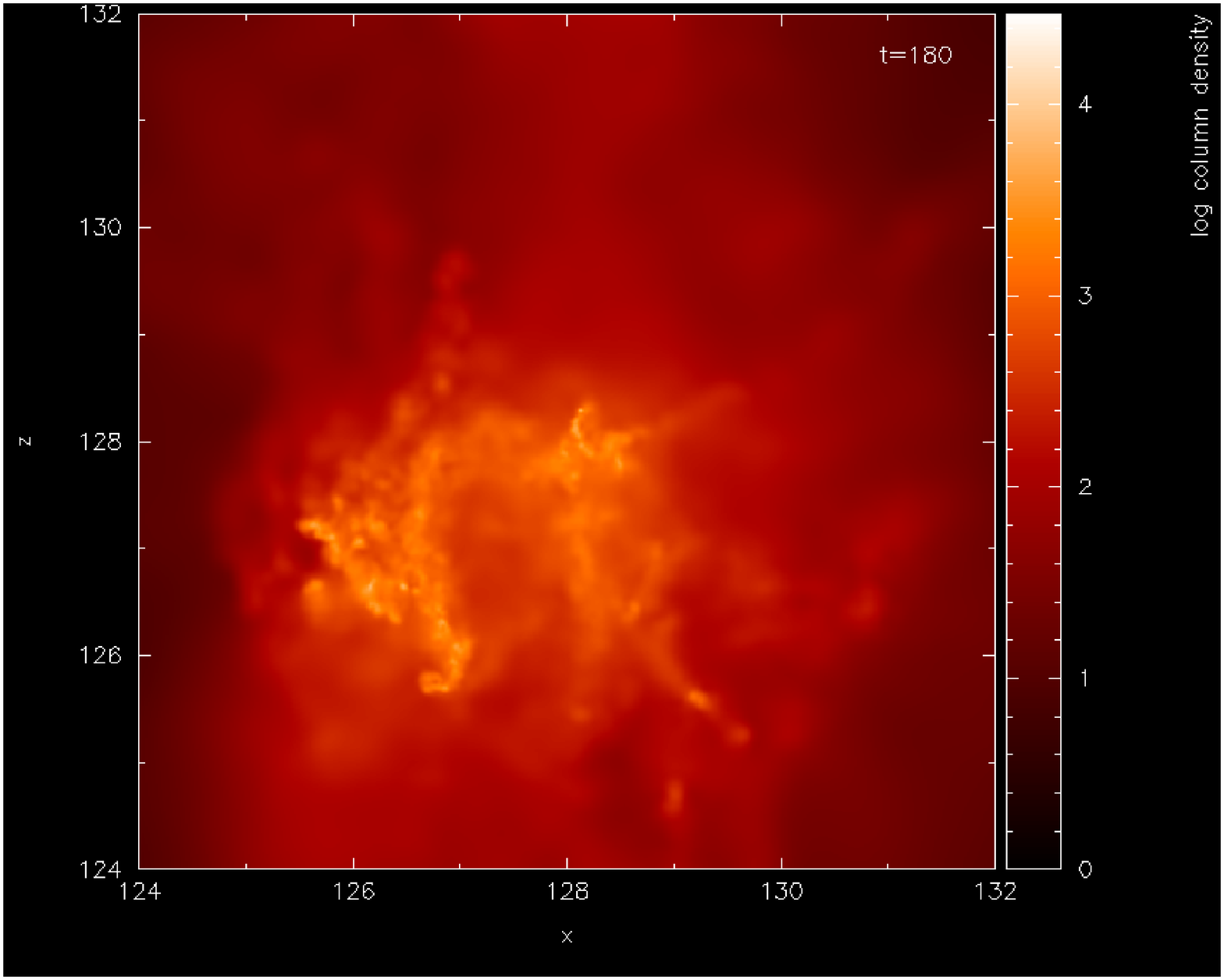}{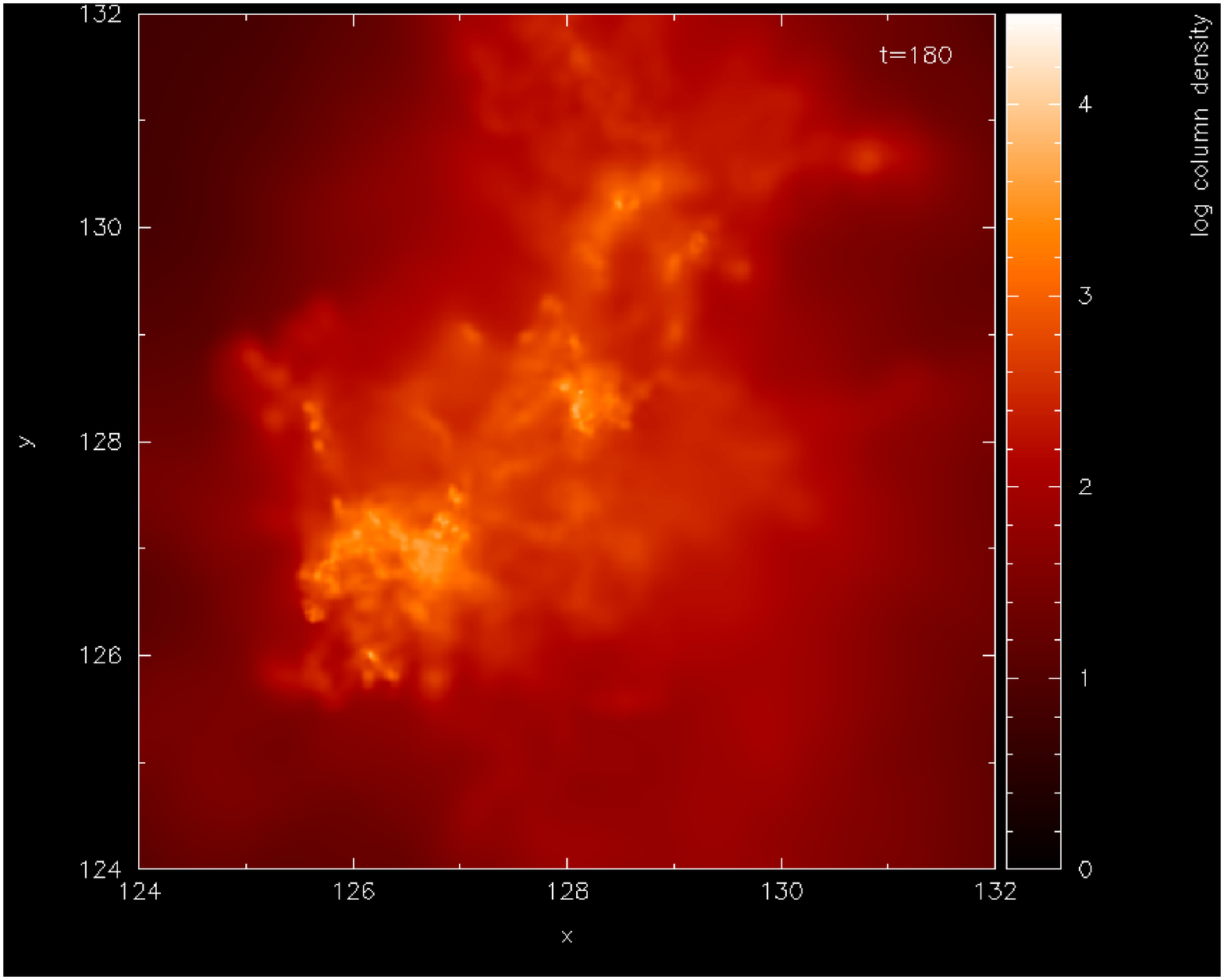}
\caption{Two views of the central 8-pc cubic region. {\it Left panel:}
Column density integrated along the $y$ direction. {\it Right panel:}
Column density integrated along the $z$ direction. The electronic
version of these images show the evolution of these regions from $t=22.6$
to 25.2 Myr.}
\label{fig:cen8pc_images}
\end{figure}

Figure \ref{fig:core1_images} shows a column density map of the LMR at
$t=19.1$ Myr. The image is integrated over 16 pc in the
$x$-direction. In the electronic version of the paper, this figure
corresponds to an animation showing
the evolution of this region from $t = 16.6$ to 19.9 Myr. The dots show
the newly formed sink particles. Figure \ref{fig:cen8pc_images} shows
images of the central 8 parsecs of the simulation (the HMR), at $t=24.2$
Myr, viewed edge-on ({\it left}) and face-on ({\it right}). In the
electronic version, this figure shows animations of the same region from
$t = 22.6$ to 25.2 Myr for both views. In the images and animations of
Fig.\ \ref{fig:cen8pc_images} we do not show the sink particles because
a large number of previously formed particles is already present in the
region, and it is not easy to see the particles formed there.


\section{Physical conditions of the ``low''- and ``high''-mass regions}
\label{sec:phys_cond_lo_hi}

\begin{figure}
\plottwo{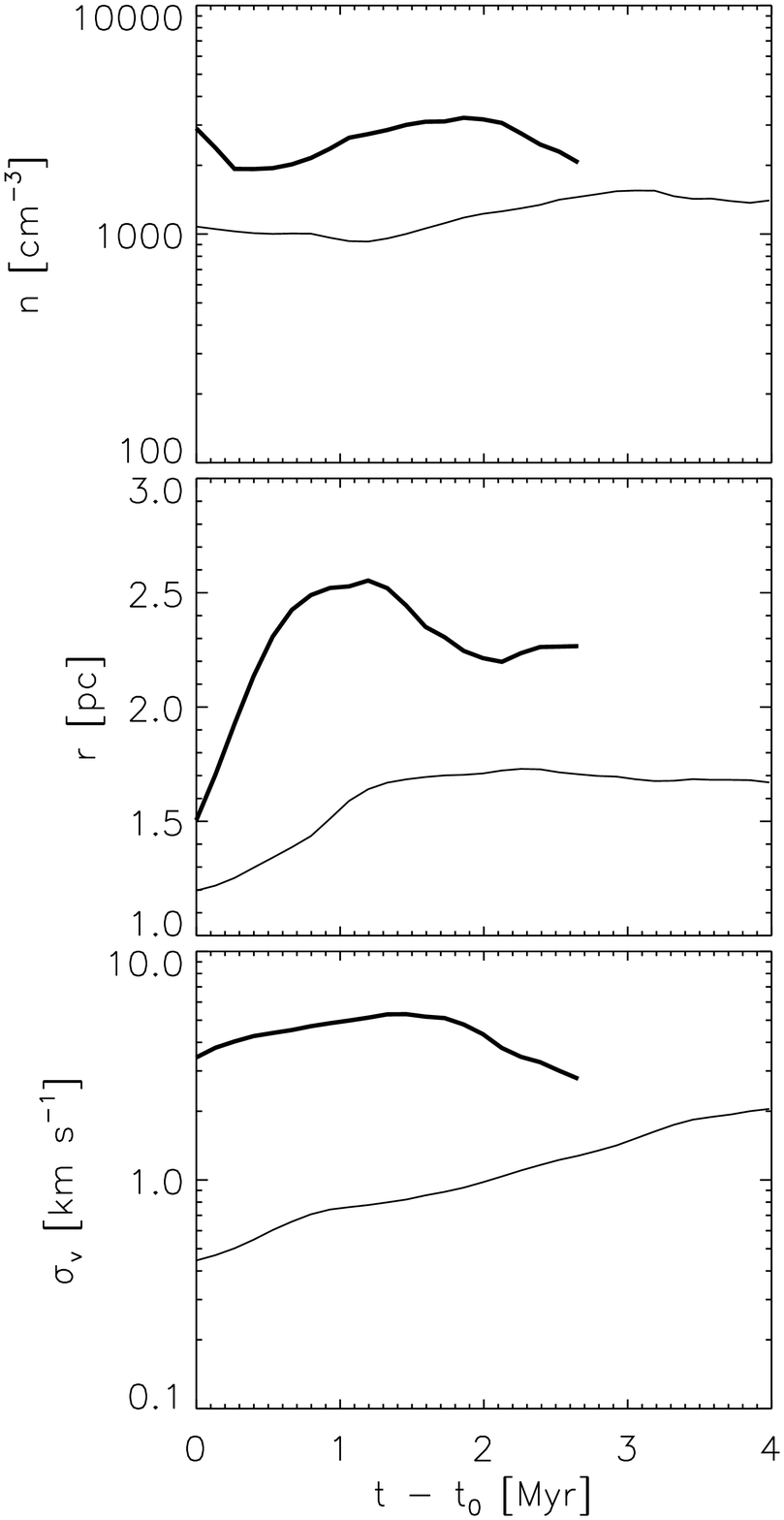} {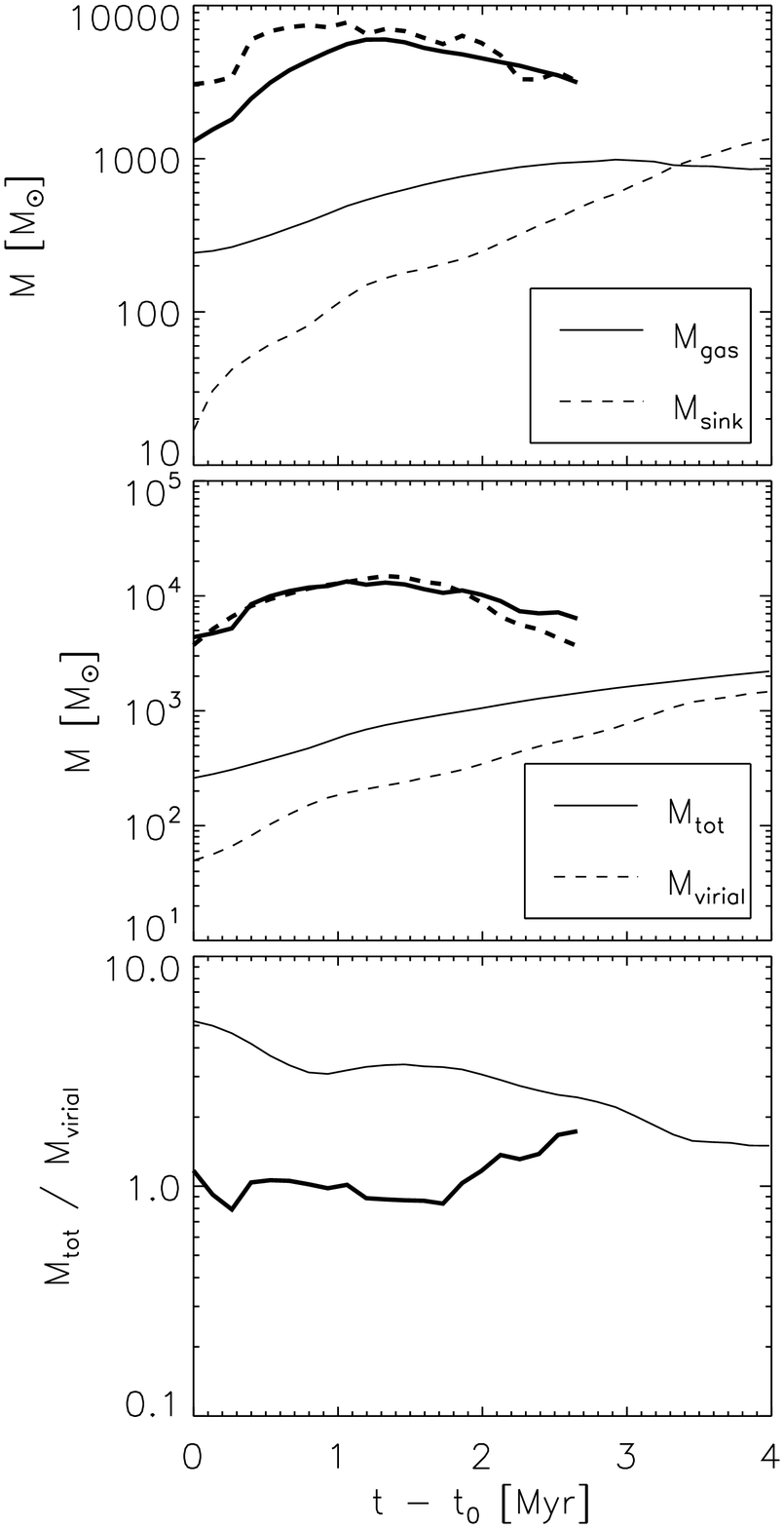}
\caption{Evolution of the physical properties of the dense gas ($n > 
\nth = 500 \pcc$) in the two star-forming 
regions anlyzed in the simulation. The {\it thick lines} refer to the
``Central Cloud'' (CC), and the {\it thin lines} refer to ``Cloud 1''
(C1). {\it Left column:} Evolution of the mean density ({\it top
panel}), size ({\it middle panel}), and velocity dispersion ({\it bottom
panel}). {\it Right panel:} Evolution of the dense gas mass and sink
mass ({\it top panel}); evolution of the total (dense gas+sinks) mass
and the virial mass ({\it middle panel}); evolution of the ratio of
total to virial mass ({\it bottom panel}). Note that the starting times
$t_0$ for the evolutionary plots are different for the two regions,
being $t_0 = 17.27$ Myr for C1 and $t_0 = 22.58$ Myr for the CC.}
\label{fig:lo_hi_comparo}
\end{figure}

We begin our comparison study between the two regions by measuring the
evolution of physical and star-forming properties of the ``clouds'' in
these regions.  To this end, we interpolate the SPH data of the regions
shown in Figs.\
\ref{fig:core1_images} and \ref{fig:cen8pc_images} onto a $256^3$ grid
in order to manipulate the data with the IDL software. We
then define the clouds as connected sets of pixels with densities
above $\nth = 500 \pcc$. We refer to the resulting clouds as ``Cloud 1''
(C1) in the LMR, and ``the Central Cloud'' (CC) in the HMR. In Fig.\
\ref{fig:lo_hi_comparo} we compare the evolution of the physical
properties of the two clouds over a few Myr. Note that the starting
times ($t_0$) for the evolutionary plots are different in the two cases,
being $t_0 = 17.27$ Myr for C1 and $t_0 = 22.58$ Myr for the CC. The
{\it thick lines} correspond to the CC, and the thin lines correspond to
C1. The {\it left column} shows the evolution of the density, size and
velocity dispersion for the two clouds, where the size is calculated as
$r =
\left(3 V/ 4\pi\right)^{1/3}$, $V$ being the volume of the cloud.

From these figures we see that the CC clearly has 
larger density, size, velocity dispersion, and mass than C1, at
least during the time spans shown in the figures (C1 later acquires
more mass and becomes more similar to the CC). Although the
density of the CC is only a factor of 2 larger than that of C1
on average, we see that its size ranges between half and one order of
magnitude larger than that of C1, and similarly for its velocity
dispersion, which is on the order of a few $\kms$, as is the case for
HMRs \citep[e.g.,][]{Beuther_etal07}. Similarly, the CC's
mass ranges in the thousands of solar masses, while that of C1
initially ranges in the hundreds, although, as mentioned above, this
region is growing in time, and approaching the conditions of the
CC at late times. Note that the gas mass in the CC
is initially less than the sink mass because we
include the mass of pre-existing sinks in the region, and not
only the sinks formed by the CC. We do this because we next compute
with the virial mass of the system, which must be compared with the
total gravitational mass, including the gas and all sinks in the region.

We compute the virial mass of each region applying the standard formula 
\begin{equation}
M_{\rm vir}\equiv 210  \left(\frac{R} {{\rm pc}}\right)
\left(\frac{\Delta v_{\rm eff}}{{\rm km\ s^{-1}}}\right)^2 M_\odot.
\label{eq:virial_mass}
\end{equation}
\citep[see, e.g.,][]{Caselli_etal02, Tachihara_etal02,
Klessen_etal05}. 
It is noteworthy that the CC, even though it is in an extremely dynamic
infalling state, has a ratio of mass to virial mass $M/\MJ$ of almost
exactly unity during most of its evolution. This is precisely what is
expected for an object undergoing gravitational contraction, and has
been observed previously in numerical simulations \citep[e.g.,][]{KB00,
KB01, VS_etal07}. Instead, this ratio is completely arbitrary in the
case of clouds or clumps formed purely by turbulent compressions.

Cloud 1, on the other hand, starts with a ratio of nearly 5,
but it steadily approaches unity as it accelerates in its contraction. Note
that such values of this ratio are not uncommon in observed low-mass clouds
\citep[see, e.g.,][]{Tachihara_etal02, Morata_etal05}. Presumably, the
relatively large initial values of the ratio indicate
that at its initial states, the cloud is only beginning to decouple from
the global flow, and starting its own collapse.

\begin{figure}
\plotone{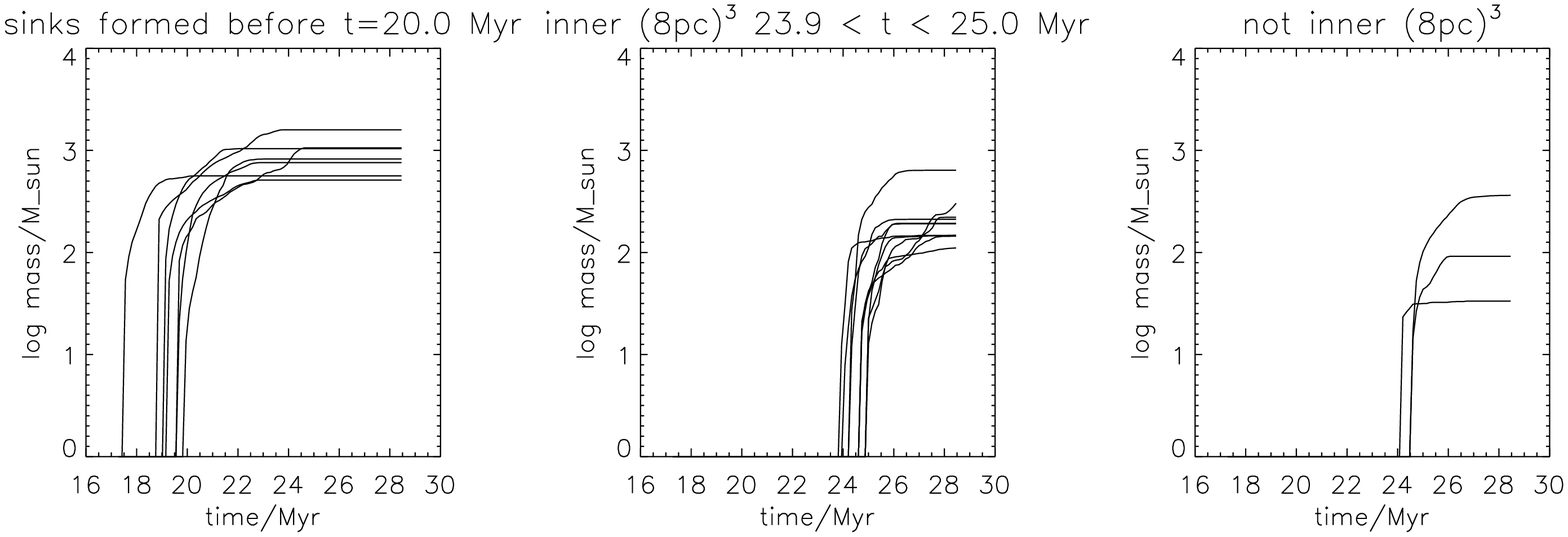}
\caption{Evolution of the most massive
sinks formed in the simulation up to $t=20$ Myr ({\it left panel}), the
most massive sinks formed in the central 8-pc region for $23.9 < t < 25$
Myr ({\it middle panel}), and those formed outside of the central 8-pc
region during the same time interval ({\it right panel}).}
\label{fig:sink_mass_evol}
\end{figure}

We furthermore measure the {\it sink} formation rate (SiFR) in both clouds,
defined as
\begin{equation}
{\rm SiFR} = \frac{\Delta \Msink}{\Delta t},
\label{eq:SFR}
\end{equation}
where $\Delta \Msink$ is the increment in sink mass in the region during
the time interval $\Delta t = 0.13$ Myr between successive data
dumps. Analogously, we define the ``specific SiFR'' ({\it sifr}) as
\begin{equation}
{\it sifr} = \frac{{\rm SiFR}}{\Mgas},
\label{eq:specific_SFR}
\end{equation}
where $\Mgas$ is the instantaneous dense gas mass in the region, 
again defined as the mass of gas with densities $n > 500 \pcc$. Note
that, for simplicity, we
perform these measurements directly on the SPH data, rather than on
their interpolation on a grid. Note also that we use the increment
in the mass of the sinks rather than in the number of sinks because, as
mentioned in \S \ref{sec:model}, our
sinks are not sufficiently resolved in mass, and furthermore
they continue to accrete mass after they form, as shown in Fig.\
\ref{fig:sink_mass_evol}. This figure shows the mass evolution of the
most massive sinks formed in the simulation up to $t=20$ Myr ({\it left
panel}), the most massive sinks formed in the central 8-pc region for
$23.9 < t < 25$ Myr ({\it middle panel}), and those formed outside of
the central 8-pc region during the same time interval ({\it right
panel}).

\begin{figure}
\plotone{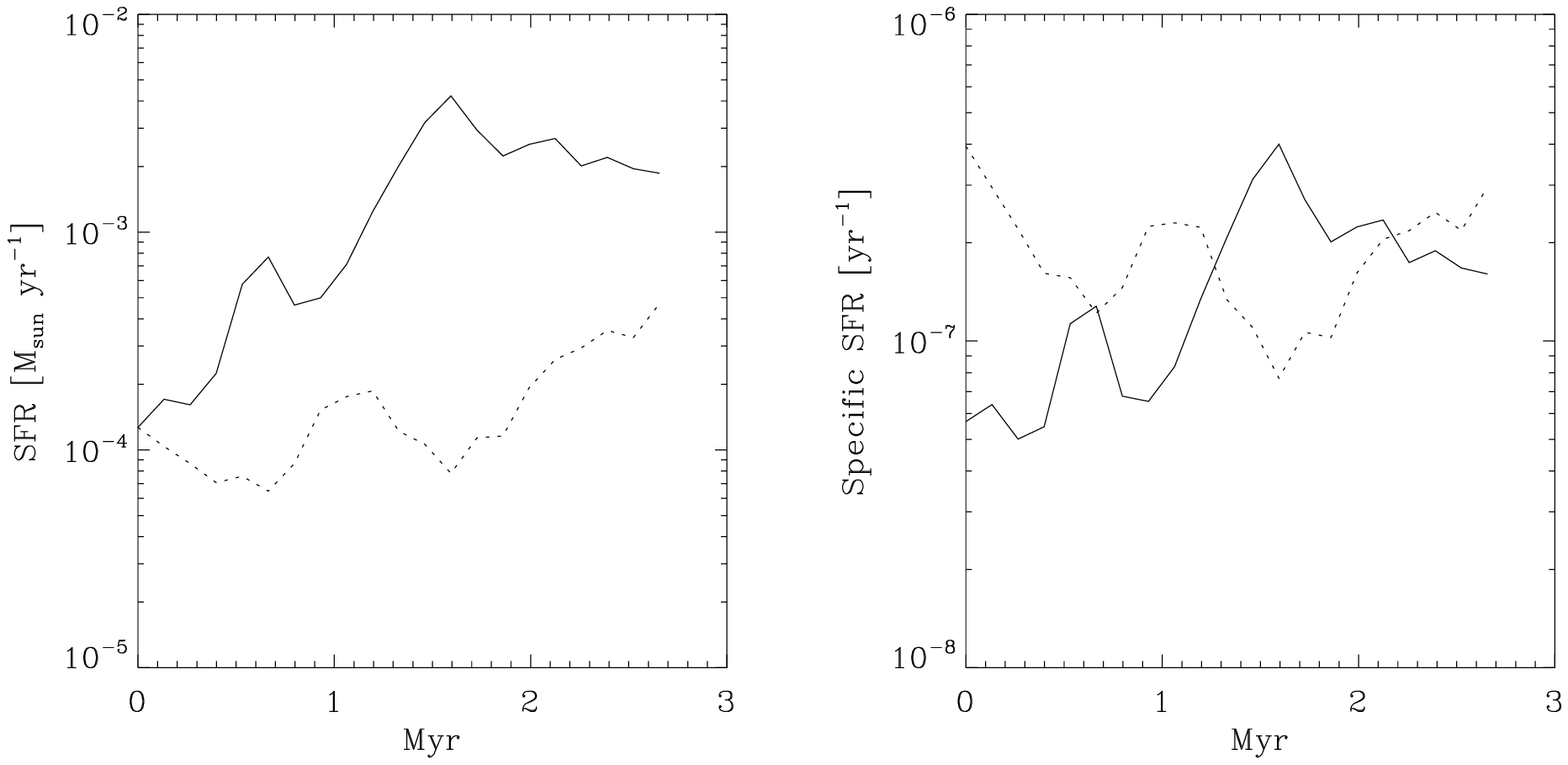}
\caption{Evolution of the SiFR and the specific SiFR ({\it sifr}) for
Cloud 1 (C1, {\it dotted lines}) and the Central Cloud (CC, {\it solid
lines}).}
\label{fig:SFRs}
\end{figure}

Both the SiFR and the {\it sifr} are shown in Fig.\ \ref{fig:SFRs}. It is
clearly seen that again the SiFR of the CC is roughly one order of
magnitude larger than that of C1. However, it is seen that the {\it
specific} star formation rate is similar for both clouds, suggesting
that the larger SiFR of the CC is due exclusively to the larger amount of
mass available for collapse in this cloud.
In any case, if one assumes a universal IMF, then the larger SiFR of the
CC implies that it has a ten-fold larger likelihood of forming massive
stars, in correspondence with its more massive and violent status.

\section{Comparison with observations}
\label{sec:comparison_obs}

In this section we compare the mean physical properties of the massive
clump in the CC with those of a typical massive-star forming clump in
the Orion Molecular Cloud, namely the well-known clump OMC-1. Next we
compare the statistical distributions of masses, sizes and densities of
the dense cores in the Cygnus-X North region, reported by
\citet{Motte_etal07} with the corresponding structures within the CC.

It is important to note that, since we are making these comparisons against
data from two different molecular clouds, we are making the implicit
assumption that the massive-star forming clumps in GMCs are similar in
different clouds, constituting a general class of objects. This is, however, a
reasonable assumption, since these clumps are often discussed in general
terms, regardless of the GMC to which they belong \citep[see, e.g., the
discussion in][sec.\ 2.1]{Beuther_etal07}.


\subsection{Physical properties and star formation rates}
\label{sec:phys_props_SFRs}

We begin by considering a typical HMR, namely OMC-1, the gas behind the
Orion Nebula Cluster (ONC) in the Orion Molecular Cloud (OMC). According to
\citet{Bally_etal87}, the total gas mass in the clump known as OMC-1 is $\sim
2200~\Msun$ and the typical observed (i.e., 1D) rms velocity is $\sim 2
\kms$, with an angular size $\Delta \theta \sim \sqrt{108}$ arcmin (see
the first entry in their Table 1). Assuming a distance of 400--500 pc to
OMC-1 \citep{Genzel_etal81, Sandstrom_etal07}, this value of $\Delta
\theta$ implies a physical size $\sim 1.2$--1.5 
pc for this cloud. Thus, the mean density in OMC-1 is $\sim 2200 \Msun /
(1.35~{\rm pc})^3 \sim 900~\Msun {\rm pc}^{-3} \sim 1.54 \times 10^4
\pcc$, where for the last term we have assumed a mean particle mass of
2.36. 

The above numbers can be compared with the physical conditions in the CC
in our simulation. To do this, we need to select the gas within this
region with similar mean density. We have found that this is
accomplished by using a density threshold of $4500 \pcc$ to define the
region, which gives us a mean density of precisely $1.54 \times 10^4 \pcc$. 

Next, we apply a clump-finding algorithm based on
identifying connected sets of grid points whose densities are above this
threshold, allowing us to identify the most massive clump within the
CC.  This region can be considered the
equivalent to what is often referred to as a ``cluster-forming core''
\citep{LL03}. However, we refer to it as a ``clump''. Once the clump is
defined, we can straightforwardly measure its physical properties. We
find a gas mass $M_{\rm gas} \sim 3015~\Msun$, a size $r \sim 1.14$ pc
(computed through the formula $r = (3V/4\pi)^ {1/3}$, where $V$ is the
volume of the clump), and a three-dimensional velocity dispersion
$\sigma \sim 4.74 \kms$, implying a one-dimensional velocity dispersion
of $\sim 2.74 \kms$. We thus find that there is very good agreement
between the physical conditions in OMC-1 and those of an equivalent
region within the CC in our simulation.

Moreover, from Fig.\ \ref{fig:SFRs} we see that the CC is characterized
by a ${\rm SiFR} \sim 10^{-3} \Msun$ yr$^{-1}$ on average. Noting that
our density threshold for the formation of a sink, of $3.2 \times 10^6
\pcc$, is comparable to the density of a dense core, we should consider
that the {\it star} formation efficiency (SFE) within our sinks is less
than 100\%. Assuming an SFE $\sim 30$--50\%, we estimate a {\it star}
formation rate (SFR) $\sim 3$--$5 \times 10^{-4} \Msun$ yr$^{-1}$. On
the other hand, \citet{Tobin_etal09} report on 1613 stars in the
ONC. Taking this number as a proxy for the total stellar production of
this region, and a mean stellar mass of $0.3~\Msun$ \citep{HC00}, this
implies a total stellar mass of $\sim 500 \Msun$. The estimated age
spread of the cluster is $\lesssim 2$ Myr \citep{Hillenbrand97}. We thus
infer SFR $\gtrsim 2.5 \times 10^{-4} \Msun$ yr$^{-1}$, in very good
agreement with that of our CC.

With these data, we can also calculate the SFE per free-fall time,
\sfeff.\footnote{This quantity is frequently referred to as SFR per
free-fall time in the literature. However, it is actually the {\it
specific} SFR, integrated over a free-fall time \citep[e.g.][] {KM05,
KT07}.} Specifically, writing 
\begin{equation}
\hbox{\sfeff} = \frac{M_{\rm stars}(\tff)}{M_{\rm gas} + M_{\rm stars}(\tff)},
\label{eq:def_SFEff}
\end{equation}
where $M_{\rm stars}(\tff) \approx \hbox{SFR} \times \tff$, with SFR being the
instantaneous star formation rate, estimated above, and $\tff$ the
free-fall time. At the mean density of $1.54 \times 10^4 \pcc$, and a
temperature of 10 K, we have $\tff \approx 0.27$ Myr. We thus find
\begin{equation}
\hbox{\sfeff} \approx \frac{135 \Msun}{3150 \Msun} = 0.043.
\label{eq:SFEff_CC}
\end{equation}
Interestingly, this value is fully consistent with that obtained by
\citet{KT07} ($\hbox{\sfeff} \sim$ 0.03--0.09) for the ONC. We discuss this
result further in \S \ref{sec:sfeff}.

Finally, we can also perform a comparison between the CC and OMC-1 at
the level of the {\it core} formation rate. \citet{Ikeda_etal07} report
a total mass of dense cores in the OMC-1 region of $\sim
800~\Msun$. Assuming that these have typical lifetimes between 2 and $5
\times 10^5$ yr \citep[e.g., ][]{Hatchell_etal07}, we find a core
formation rate (CFR) between 1.6 and $4 \times 10^{3} \Msun$ Myr$^{-1}$,
which again compares very well to the mean value of the SiFR mentioned
above. Note that the estimated CFR for OMC-1 is larger than the
estimated SFR in the ONC by a factor of a few, a reasonable result if
the efficiency of star formation within the cores is smaller than unity,
and consistent with the factor we have used above for estimating the SFR
from the SiRF.

We conclude that the physical conditions and SFR in a well-known
massive-star forming clump agree to within less than factors of 2 with
the corresponding features in the CC of our simulation.


\subsection{Statistics of cores in the Central Cloud}
\label{sec:core_stats_ctr}


As a further test of the similarity of the CC with massive-star forming
clumps, in this section we compare the statistics of the dense cores
within the CC with those reported by \citet{Motte_etal07} for the Cygnus
X molecular complex. These authors have performed a 1.2 mm continuum
unbiased survey of the massive-star forming clumps within the Cygnus-X
complex, and reported the masses, sizes and densities of the cores
within the clumps, and of the clumps themselves. Specifically, they
compiled a large dataset of 129 massive dense cores, of sizes $\sim 0.1$
pc, masses $M_{\rm 1.2~mm} = 4$--950 $\Msun$, and densities $n \sim 10^5
\pcc$, listed in Table 1 of that paper. These data are very well suited
for comparison with the corresponding structures in the CC of our
simulation.  In order to perform the comparison, we identify dense cores
within the gridded version of the CC in our simulation (cf.\ Fig.\
\ref{fig:cen8pc_images}) at time $t = 23.9$ Myr, at which the global
collpase of the entire cloud complex in the simulation is being
completed (i.e., the clouds that had initially formed at the peripheral
ring in the simulation reach the center at this time).

To define the cores, we apply the same
clump-finding algorithm used in \S \ref{sec:phys_props_SFRs}, but in
this case we use it iteratively and with higher threshold densities to
define an ensemble of cores. Specifically, we consider a set of four
thresholds, at $n = 10^4$, $3 \times 10^4$, $10^5$, and $3 \times 10^5
\pcc$, and include all of the resulting data points into a single
dataset. We do this in order to have a sufficiently large sample, and
to allow for a wider dynamic range in the physical properties of the
cores, since our previous experience shows that objects selected with a
given threshold have mean densities that differ by less than an order of
magnitude from the value of the threshold \citep[e.g.,][]{VBR97}. Note
that our procedure differs from other algorithms such as CLUMPFIND
\citep{WdGB94} in that, in our case, clumps defined at a lower threshold
may contain several clumps defined at a higher threshold, while
CLUMPFIND, for example, does not identify such ``parent'' structures as
a clump. We choose this procedure for simplicity, and
because larger clumps defined at lower thresholds may be bound
structures themselves that should be taken into
account. \citet{Motte_etal07} themselves used a combination of source
extraction analysis and GAUSSCLUMPS \citep{SG90}.

\begin{figure}
\plotone{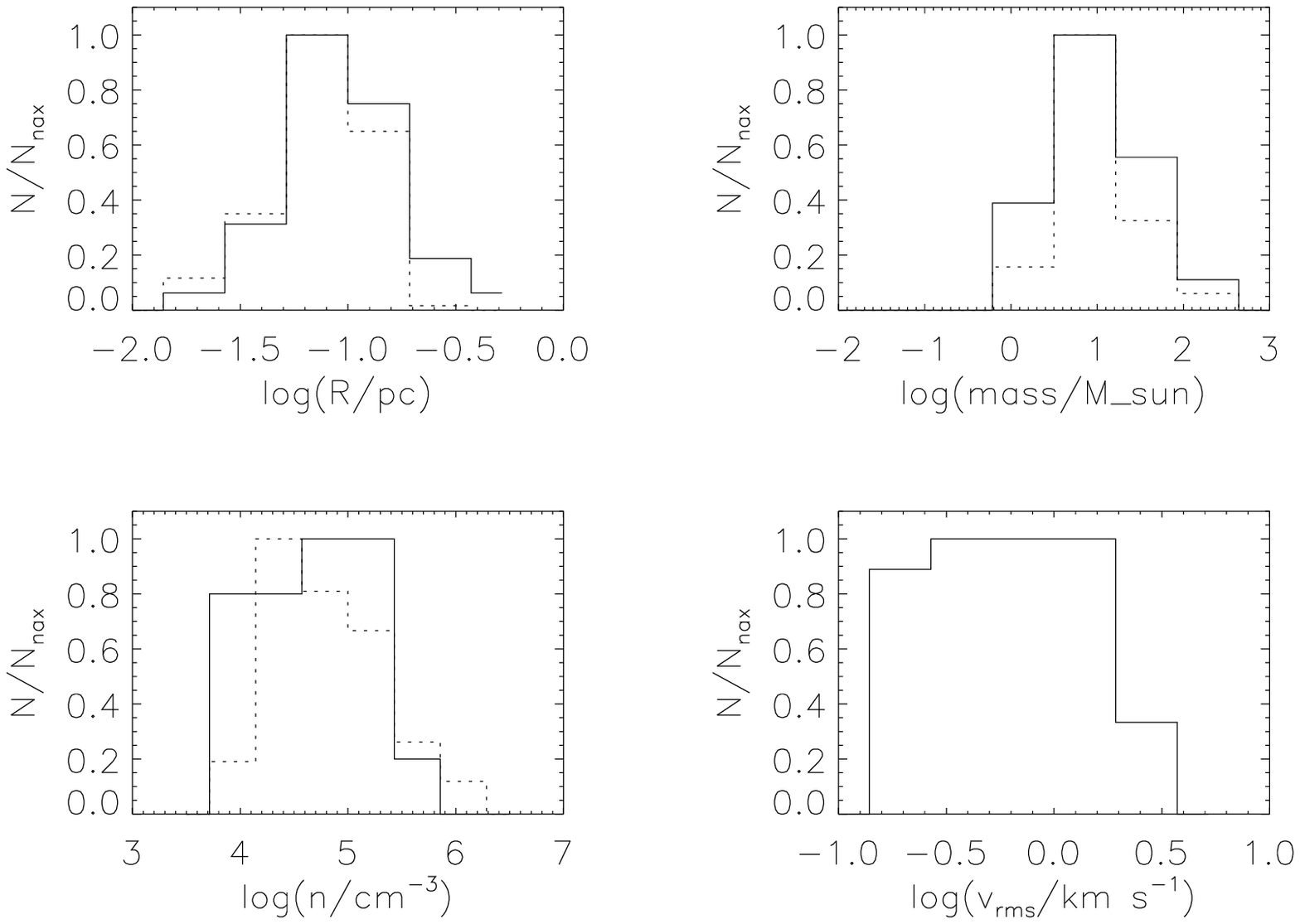}
\caption{Histograms of the size ({\it top left}), mass ({\it top
right}), mean density ({\it bottom left}) and three-dimensional velocity
dispersion ({\it bottom right}) of the cores within the Central Cloud of
the simulation ({\it solid lines}) at $t = 23.9$ Myr, at which time the
region acquires its most compressed state. The histograms are normalized
to their peak values. {\it Dotted lines}: same for the core
sample in Cygnus X reported by \citet{Motte_etal07}, except for the
velocity dispersion, since their observations were performed in the
1.2~mm continuum. The distributions are seen to match extremely well for
size, mass and mean density.}
\label{fig3}
\end{figure}

This procedure leaves us with 39 cores, 38 of which have $M > 4
M_\odot$. For each core, we measure its mass, mean density, and velocity
dispersion, and estimate its size as $r \approx (3 V/4 \pi)^{1/3}$,
where $V$ is its volume. The distributions of these properties can be
compared with those reported by \citet{Motte_etal07} for any of the
regions within the Cygnus-X complex. As a representative example, we
choose the data for the Cygnus-X North region, which contains 72 cores
(see their Table 1). We show the distributions of core properties for
the Cygnus-X North region in the histograms presented in Fig.\
\ref{fig3} by the {\it dotted} lines. Superimposed on these histograms,
the {\it solid} lines show the corresponding distributions for the 38
cores more massive than $4 M_\odot$ in the CC of our simulation. We see
that the distributions match extremely well for size, mass, and mean
density. No comparison is made for the velocity dispersions because
\citet{Motte_etal07} do not report them, but in our sample, they range
from $\sim 0.2$ to $\sim 3 \kms$.

A final comparison that can be made is at the ``clump''
level. This is analogous to the comparison we made with OMC-1, but
comparing against the mean properties of the clumps identified by
\citet{Motte_etal07}. They report the typical sizes, masses, and
densities of all clumps to be $r \sim 0.68$ pc, $M \sim 1000 M_\odot$,
and $n \sim 1.4 \times 10^4 \pcc$ respectively (see column 3 in their
Table 4). Again, this is comparable to the clump we obtain thresholding
our CC data at $\nth = 4500 \pcc$, for which we find a mean density of
$n \approx 1.54 \times 10^4 \pcc$, a size $r \approx
1.14$ pc, and a mass of $3015~\Msun$ (cf.\ \S
\ref{sec:phys_props_SFRs}). We see that our clump is somewhat more
massive and extended than \citet{Motte_etal07}'s mean clump, but still
within factors of 2--3 of the latter.


\section{Discussion and caveats} \label{sec:discussion}

\subsection{The star formation efficiency per free-fall time} \label{sec:sfeff}

The results from \S \ref{sec:phys_props_SFRs} have a number of
interesting implications. We note that the SF {\it rates} in both the CC
in our simulation and in OMC-1 in Orion are relatively large, $\sim 3
\times 10^{-4} \Msun$ yr$^{-1}$. This can be translated into a {\it
specific} SFR (or inverse SF, or gas depletion, timescale, $\tsfinv$),
given by
\begin{equation}
\tsfinv \approx \frac{\hbox{SFR}} {\Mgas + \Mstar},
\end{equation}
where $\Mgas$ is the gas mass in the cloud and $\Mstar$ is the stellar
mass. We obtain $\tsf \sim 10$ Myr for both the CC and OMC-1. This
differs strongly from the much longer timescales $\tsf \sim 2$ Gyr for
extragalactic GMCs \citep{Bigiel_etal08}, and $\tsf \sim 300$ Myr for
Galactic GMCs \citep{MW97}. Said another way, if one were to extrapolate
(linearly with mass) the SFR of these clumps (the CC or OMC-1) to the
entire molecular gas mass of the Galaxy of $10^9 \Msun$, one would find
a total Galactic SFR of $\sim 3\times 10^{-4} \Msun$ yr$^{-1} \times
10^9 \Msun / 3000 \Msun = 100~\Msun$ yr$^{-1}$, or roughly 50--100 times
larger than the observed SFR. This is the well known \citet{ZP74}
conundrum. However, also in section \S \ref{sec:phys_props_SFRs}, we
showed that the \sfeff\ of both regions is small, and fully consistent
with the estimate of \citet{KT07} for ONC, and the general result of a
low \sfeff\ for the cores in the survey by \citet{Evans_etal09}.

These results then raise two important questions. The first is how to
reconcile the much larger specific SFR of regions like the CC or OMC-1
with that of entire molecular clouds, or even the total molecular gas
content of the Galaxy. The large disparity in $\tsfinv$ between
individual cluster-forming clumps and large-scale objects like GMCs
can be understood if star formation is a {\it spatially intermittent}
phenomenon. That is, it does not occur everywhere in the molecular gas,
but only at special locations, characterized by the highest
densities. Thus, the small fraction of the mass that {\it is} forming
stars must have a much larger specific SFR than the global average.
This is analogous to the intermittency of the energy dissipation rate in
incompressible turbulent flows, which is known to occur only at
scattered locations, where the velocity gradient is maximized by vortex
stretching \citep[see, e.g., ][]{Frisch95}.

The second question is: how can such an actively star forming region have
such a small value of the \sfeff. The theory by \citet{KM05} originally
derived such low values for the \sfeff\ for clumps in near virial
equilibrium supported by turbulence \citep[similarly to the underlying
assumption of the model by][] {MT03}. However, the CC in our simulation
is far from being in equilibrium, as shown in the animations
corresponding to Fig.\ \ref{fig:cen8pc_images}.

\begin{figure}
\plotone{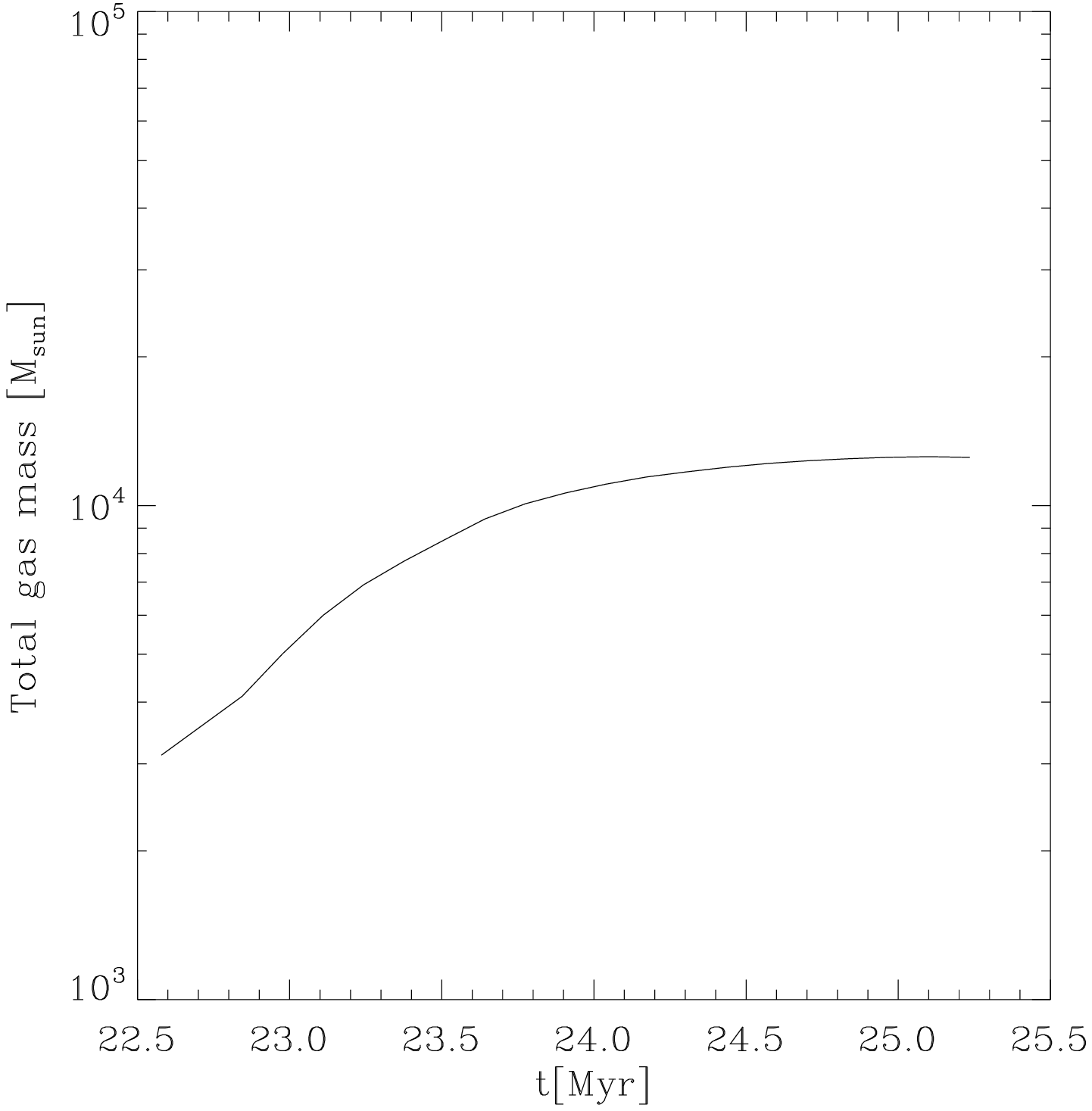}
\caption{Evolution of the total gas mass in the central 8-pc box of the
simulation. The total mass in this region is seen to increase from $\sim
4 \times 10^3 \Msun$ to $\sim 1.2 \times 10^4 \Msun$ in 3 Myr,
corresponding to a mean accretion rate of $2.7 \times 10^{-3} \Msun$
yr$^{-1}$.} 
\label{fig:mass_tot_evol}
\end{figure}

The origin of the small \sfeff\ value for the CC in our simulation can
be understood by rewriting eq.\ (\ref{eq:def_SFEff}) as
\begin{equation}
\hbox{\sfeff} = \frac{\hbox{SFR}}{M_{\rm tot}}~ \tff,
\label{eq:def_SFEff_2}
\end{equation}
where $M_{\rm tot}$ is the total mass (gas + stars) in the cloud. From
this expression we see that a low value of the \sfeff\ can be obtained
through a large cloud mass and/or a high density (via $\tff$), besides
the obvious possibility of a small SFR. This appears to be the case of
the CC, which is rapidly increasing its mass through accretion of the
surrounding gas at a rate that apparently overwhelms its consumption
rate by SF. Indeed, Fig.\ \ref{fig:mass_tot_evol} shows the evolution of
the total gas mass in the central 8-pc box of the simulation in which the CC
is contained. The gas mass is seen to increase from $\sim 4000 \Msun$ to $\sim
12000 \Msun$ in 3 Myr, corresponding to a mean accretion rate onto the
region of $\sim 2.7 \times 10^{-3} \Msun$ yr$^{-1}$, or roughly an order
of magnitude larger than the SFR in the CC.

Moreover, because of the ram pressure (or, equivalently, weight) of the
accreting material, the density is also larger than what would be
expected if the clump were isolated and in hydrostatic equilibrium.
That is, the actual collapse timescale of a region such as our CC and
its surroundings is {\it longer} than the free-fall time of the central
dense core alone, because the mean density of the combined system is
lower than that of the core, and the material is continuously being
replenished at the central regions. Thus, if the infalling material is unseen
for any reason (e.g., thresholding, background substraction, termination
of the tracer excitation, etc.), then the central core will appear to
have a shorter free-fall time than the actual timescale of the
collapse. This can be incorrectly interpreted as an isolated,
equilibrium core with a lifetime much longer than its local free-fall
time and thus with an SFR much slower than that given by the free-fall
rate.

Finally, it is worth noting that this scenario then naturally explains
the large final efficiencies of cluster-forming clumps
\citep[$\sim 30$--50\%; e.g.,][]{LL03}, without having to invoke
lifetimes for these clumps that are much larger than their free-fall times.
Also, in this accreting scenario, we see that {\it a low local
value of the \sfeff\ does not guarantee a globally low SFR.} Our results
show that a region can have a large SFR and still have a low \sfeff. The
attainment of a globally low SFR seems to require a highly scattered and
intermittent occurrence of the star-forming regions.

\subsection{Neglect of stellar feedback}
\label{sec:feedback} 

The simulation we have analyzed in this paper neglects two important
processes existing in real molecular clouds, namely stellar feedback and
magnetic fields. Concerning the former, it is generally believed that it
is a key mechanism in keeping a low final SFE in molecular clouds and
their clumps. However, at present it is still unclear whether this
self-regulation of SF occurs abruptly, after a brief
period of a high SFR, dispersing the star-forming region and quenching
SF altogether locally \citep[the so-called ``rapid SF'' scenario;
e.g.,][] {Whit79, Elm83, Elm00, Elm07, Larson87, FST94, HBB01}, or
whether it occurs continuously over longer times,
maintaining the region in a near-equilibrium state, allowing it to last
for several free-fall times, while maintaining a low SFR \citep[the
so-called ``slow SF'' scenario; e.g.][]{KM05, KMM06, LN06, KT07}.

The behavior of our simulation is consistent with the ``rapid SF''
scenario, in which a given region forms stars at a high rate over a
short period of time and then their feedback disrupts the cloud,
possibly assembling a new star forming region at a new location, through
the ``collect and collapse'' mechanism \citep{EL77, Elm07}. Since the
large complex and its fragments in our simulation begin their
gravitational contraction several Myr before the first star-forming
collapse events occur (cf.\ \S \ref{sec:evolution}), it is natural to
expect that the initial stages of SF in each region will occur in a
gravitationally contracting environment, and to be characterized by a
large SFR, until enough stars have been formed that they can strongly
affect, and possibly disrupt their parent clump \citep[e.g.,][] {Elm83,
Larson87, FST94}. 

Under this scenario, our simulation and the star-forming regions formed
within it represent the assembly and early star-forming epochs of a
region, although their long-time star formation efficiencies are too large
($\sim 60$--70\% after $\sim 10$ Myr; cf.\ Paper I) because they lack
the subsequent disrupting effect of the stellar feedback on the clump in
which the stars form, and so SF continues unimpeded in each region until
the gas is exhausted.  However, this is a shortcoming that is only
expected to become important later into the evolution of each region,
when the accumulated number of stars becomes large enough to exert a
strong influence on the cloud. As mentioned in
\S \ref{sec:evolution}, in Paper I we applied a prescription by
\citet{FST94} to estimate this time, which we found to be $\sim 3$ Myr
after the onset of SF, although new simulations including stellar
feedback\citep[][\VS\ et al., in prep.]{Dale_etal05, DB08,
Peters_etal09} suggest that the effect of feedback is 
more complicated than that, and should be applied clump by clump. But,
in any case, our simulation should be accurate before the time of
disruption by feedback, for each new star-forming region formed.

On the other hand, the evolution of our simulation is inconsistent with
the ``slow'' scenario of star formation \citep[e.g.,][] {KM05, KT07}, in
which clouds are assumed to be maintained in quasi-static equilibrium by
the turbulence injected by stellar sources, as modeled by
\citet[e.g.,][]{Matzner02, KMM06, LN06}. Our simulation, although
incapable of following the stages in which SF interacts strongly with
the infall process onto the clouds, suggests that they should be already
contracting in general by the time they begin forming stars, rather than
being in a quasi-hydrostatic state.  Besides, the studies suggesting
such a hydrostatic state have been performed either in closed numerical
boxes or with simplified geometries, so that the disruption of the cloud
by its stars has been made overly difficult. Moreover, it appears
unlikely that the effect of the stellar energy injection, which occurs
at small scales, can somehow organize itself to produce the large-scale,
dipolar character of the principal velocity component in the clouds
\citep{HeBr07}. Simulations in which the cloud is embedded in its
diffuse medium and in full three-dimensional geometry, such as the one
utilized here, but including the effects of stellar feedback, are
necessary to properly address the issue (\VS\ et al., in prep.)

Our results are consistent with recent
suggestions, based on comparisons between simulations and observations,
that molecular clouds \citep[e.g.,][]{HB07} and clumps
\citep[e.g.][]{PHA07} may be in a state of gravitational collapse. If
confirmed, these suggestions point towards a return to the original
suggestion by \citet{GK74} that the observed linewidths in molecular
clouds are due primarily to gravitational contraction. This suggestion
was dismissed by \citet{ZP74} through the argument that this would imply
a much larger average star formation rate in the Galaxy than observed.
However, in the scenario of rapid SF the instantaneously high rates of
SF are not a problem, if the star-forming clumps are soon disrupted by the
stellar feedback before their entire mass is consumed by SF. 

\subsection{Neglect of magnetic fields}
\label{sec:field} 

Our simulation also neglects the influence of magnetic fields, and as
such it should be considered as representative of the evolution of the
supercritical parts of the clouds, which are the ones that can undergo
global collapse. For these regions, the presence of weak but nonzero
magnetic fields can retard the collapse \citep{HMK01, OSG01, VKSB05}. In
any case, if other parts of the clouds are subcritical and are thus held
up against their self-gravity by the magnetic field, then they must
contribute only a small (or nearly null) fraction of the SF in the
Galaxy \citep{Elm07}, since it is known that most stars form in
cluster-forming regions \citep{LL03}, while a subcritical region is
expected to be a site of isolated, low-mass star formation. Of course,
if a large fraction of the cloud's mass is supported by the field, then
the global SFE for the cloud will certainly be decreased, but the
supercritical regions will still be responsible for the majority of the
SF activity of the cloud, and those parts should have similar SFEs as
our simulation. Moreover, as discussed in \S \ref{sec:sfeff}, it is
precisely this kind of structure that is necessary to have a large
fraction of the molecular mass {\it not} participating in SF, while the
remaining small fraction \citep[a few percent; see,
e.g.,][]{Kirk_etal06} can be essentially free-falling, and accounting
for most of the SF in the GMC.

\subsection{Clump properties} \label{sec:clump_props}

Although the CC has been shown to
reproduce several features of massive-star forming clumps, it does also
exhibit certain differences. Most notably, the total mass in the CC
contains roughly twice the mass in sinks than in gass at $t-t_0 = 0$
(cf.\ Fig.\ \ref{fig:lo_hi_comparo}, upper right panel). This is
contrary to the situation in the ONC-OMC-1 complex, where the gas mass
is roughly 4 times larger than the stellar mass. The excess sinks in the
CC are ``stragglers'', which were formed earlier in the peripheral ring,
but that have fallen into the large potential well of the whole complex
together with the gas. This suggests that the global collapse
responsible for the formation of the CC is somewhat too focused, perhaps
as a consequence of the extremely smooth initial conditions used in the
simulation. Indeed, experimentation with more strongly fluctuating
initial conditions in the velocity \citep{Rosas_etal09} shows that
in this case the collapse is less focused, possibly causing the
population of infalling sinks in the collapse center to be less
numerous. Nevertheless, the presence of such infalling stars may explain
the age spreads observed in regions of massive star formation
\citep{PS99, Faustini_etal09}.

Note also that, due to the presence of the infalling sinks, it is
possible for both the gas and sink mass in the upper right panel of
Fig.\ \ref{fig:lo_hi_comparo} to decrease simultaneously. This is a
consequence of the gas being consumed to form new sinks, but with the
SiFR not being able to compensate for the departure of the infalling
sinks which, contrary to the situation of the gas, simply pass through
the region, while the gas shocks and stagnates there.

Finally, it is necessary to remark that the dense cores necessarily lie at
the limit of the scales resolved by the simulation. Thus, the velocity
dispersion within them is most certainly reduced by numerical diffusion.
However, the fact that the gravitational contraction is hierarchical,
and starts at the largest scales of our cloud, implies that the
collapsing nature of the clumps cannot be an artifact of the numerical
diffusion. 

\section{Summary and Conclusions} \label{sec:conclusions}


In this paper we have presented numerical evidence that the physical
conditions in low- and high-mass star forming regions (``clouds'') can
arise from hierarchical gravitational collapse. The former regions
arise from small-scale fragments in the collapse (which however occur
first, because they originate from local high-amplitude density
fluctuations which have a shorter free-fall time), while the latter may
appear when the global, large-scale collpase is completed. The local
collapse events do not exhaust the gas in their regions because a) the
consumption time is relatively long, $\sim 5$ Myr (cf.\ Fig.\
\ref{fig:SFRs}, {\it right panel}), and b) because the clouds continue
to accrete mass from their atomic surroundings throughout their
evolution. Thus, the isolated low-mass clouds eventually collide to form
a high-mass complex. In this scenario, velocity dispersions are
caused primarily by infall motions rather than random turbulence, but
its hierarchical nature may explain why massive cores tend to have
larger velocity dispersions than low-mass ones at the same size
\citep[see, e.g., Fig. 10 of][]{GL99}, since the former only begin
contracting when they decouple from the global flow, while the latter
have been accelerating in their contraction for a longer time.

The evidence we presented consisted of two parts. First, we analyzed the
evolution of the physical properties (mass, mean density, size, velocity
dispersion, and mass to virial mass ratio), as well as of the SFRs of
two examples of the class of objects, which we called ``Cloud 1'' (C1)
and ``the Central Cloud'' (CC). We noted that the latter, which forms as
a consequence of the large-scale collapse of the cloud complex in the
simulation, contains a clump with properties consistent with those of
cluster-forming cores within GMCs, such as a gas mass $\Mgas \sim 3000
\Msun$, a velocity dispersion $\sigma_v$ of a few $\kms$, and a star
formation rate $ {\rm SFR} \sim 3 \times 10^{-4} \Msun {\rm yr}^{-1}$, while C1
is more reminiscent of low- or intermediate-mass regions, with $\Mgas
\sim 400 \Msun$, $\sigma_v \sim 0.7 \kms$ (during its initial stages, as
it later becomes more massive), and SFR $\sim 3 \times 10^{-5} \Msun {\rm
yr}^{-1}$.

Secondly, we performed a survey of the dense cores in the massive cloud
(the central 8 parsecs of the simulation at the time the large-scale
collapse converges there), and compared their statistical properties
with those reported for the dense cores in the Cygnus X cloud complex by
\citet{Motte_etal07}, finding that the distributions of both samples
match each other very well. At the slightly larger-scale, ``clump''
level, we also found good agreement between the mean properties
reported by those authors and the ``clump'' in that region. These
results suggest that indeed the convergence of the global collapse is a
good model of HMRs.

This scenario, however, has the implication that the mechanism
responsible for the low global Galactic SFR may be the result of SF
being a spatially intermittent process, so that most of the molecular
gas mass is {\it not} forming stars at any given time in the Galaxy,
while those regions that {\it are} forming stars do so at a very high
rate --- a rate that in fact accounts for most of the SF activity in
GMCs and the Galaxy. This, in turn, implies that the global specific SFR
of the molecular gas in the Galaxy does not extrapolate linearly with
mass \citep[as assumed by, for example,][]{KT07} to estimate the
specific SFR of a single SF region.

Finally, we found, somewhat unexpectedly, that the massive-SF-like
region in our simulation exhibits a low \sfeff\ ($\sim 4$\%), in spite
of having a large SFR of $\sim 3 \times 10^{-4} \Msun$ yr$^{-1}$, and it
being the site of the large-scale collapse of the cloud formed in the
simulation, rather than a hydrostatic structure supported by
turbulent pressure. We attribute this to the high density and large accretion
rate occuring in this region as a consequence of the culmination of the
large-scale collapse. In particular, the region accretes mass at an even
higher rate than that at which it is forming stars, even though the
latter rate is large. This physical scenario calls for a new class of
theoretical models that take into account the sources and sinks for the
mass of a molecular cloud.

\acknowledgements 
We gratefully acknowledge the incisive and profound questions by an
anonymous referee, which pushed us into a much deeper understanding of
the phenomena described in this paper than we initially had.
The numerical simulation was performed in the cluster at CRyA-UNAM
acquired with CONACYT grants to E.V.-S. 36571-E and 47366-F. The
animations were produced using the SPLASH visualization tool
\citep{Price07}. We thankfully acknowledge financial support from
CONACYT grants 47366-F to E. V.-S. and 50402-F to G. C. G., UNAM-PAPIIT
grant IN110409 to J. B.-P. A.-K.J. acknowledges support by the Human
Resources and Mobility Programme of the European Community under
contract MEIF-CT-2006-039569. R.S.K.\ acknowledges financial support
from the German {\em Bundesministerium f\"{u}r Bildung und Forschung}
via the ASTRONET project STAR FORMAT (grant 05A09VHA) and from the {\em
Deutsche Forschungsgemeinschaft} (DFG) under grants no.\ KL 1358/1, KL
1358/4, KL 1359/5. R.S.K.\ furthermore is thankful for subsidies from a
Frontier grant of Heidelberg University sponsored by the German
Excellence Initiative and for support from the {\em Landesstiftung
Baden-W{\"u}rttemberg} via their program International Collaboration
II.

\clearpage

\end{document}